\documentclass[aps,prd,preprint,showpacs,showkeys,tightenlines,11pt]{revtex4}
\usepackage{epsfig}
\usepackage{color}

\def\Journal#1#2#3#4#5{{\it #1} {\bf #2}, #3 (#4)}

\def\eprint#1#2{e-print #1}

\def\PLB{{Phys. Lett.}  B}

\def\PRD{{Phys. Rev.} D}

\def\PRP{Phys. Rep.}

\def\JHEP{JHEP}

\def\IJMPA{Int. J. Mod. Phys. A}

\def\CPC{Comput. Phys. Commun.}
\def\EPJC{Eur. Phys. J. C}
\def\RMP{Rev. Mod. Phys.}

\def\ibid{\textit{ibid.}}

\begin{document}

\title{ESTIMATES FOR THE ABELIAN $Z'$ COUPLINGS FROM THE LHC DATA}
%Short title for running headings:
%Abelian $Z'$ Coupling Estimates from the LHC Data

\author{A.V.~Gulov}
\email{alexey.gulov@gmail.com}
\affiliation{Dnipropetrovsk National University, 49010 Dnipropetrovsk, Ukraine}
\author{A.A.~Kozhushko}
\email{a.kozhushko@yandex.ru}
\affiliation{Dnipropetrovsk National University, 49010 Dnipropetrovsk, Ukraine}

%\begin{history}
%\received{Day Month Year}
%\revised{Day Month Year}
%\end{history}

\begin{abstract}
In this paper, we investigate the Drell-Yan process with the intermediate heavy $Z'$ boson. We use a general approach to the Abelian $Z'$ that utilizes the renormalization-group relations between the $Z'$ couplings and allows to reduce the number of unknown $Z'$ parameters significantly. In a newly proposed strategy, we estimate the LHC-driven constraints for the $Z'$ couplings to lepton and quark vector currents. To do this, we calculate the $Z'$-related contribution in the narrow-width approximation and compare the obtained values to the experimental data presented by the ATLAS and CMS collaborations. Our method allows to estimate the values of $Z'$ couplings to the $u$ and $d$ quarks and to final-state leptons.
\end{abstract}

\pacs{PACS numbers: 11.10.Gh, 12.60.Cn, 14.70.Pw}
\keywords{$Z'$ boson; LHC bounds; hadron colliders.}

\maketitle

\section{Introduction}

Searching for new particles beyond the Standard Model
(SM) is an important part of experiments at the Large Hadron Collider.
Among the scenarios of new physics a heavy neutral vector boson
($Z'$ boson) is one of the most promising intermediate states to be detected
in hadron scattering processes in the annihilation channel. This particle resides in 
popular grand unification theories (GUTs) and other models with extended
gauge sector (see Refs. \cite{leike,Lang08,Rizzo06} for review).
Considering a couple of $Z'$ models, current experiments constrain its mass to be no less than 1-3 TeV \cite{ATLAS:2013NW,CMS:2013NW}.
At the LHC $Z'$ bosons could be discovered in the Drell-Yan process
through deviations of the cross-section from the predicted SM background.

Unfortunately, observables in experiments at hadron colliders are calculated with
significant theoretical uncertainties, that arise from the
parton model of hadrons. In this situation one can only
hope to discover the most prominent signals. This is the reason
for LHC collaborations to pay special attention to searching for narrow $Z'$ resonances.

In general, to accurately describe the $Z'$ contribution to the Drell-Yan process and to take into account the possible interference effects \cite{Boos_ea_1,Boos_ea_2}, we have to consider scattering amplitudes with intermediate virtual states.
This allows to derive few-parametric observables suitable for data fitting \cite{PankovTsytrinov:2009prd,GulovKozh:2012obs}.
But if the resonance is estimated to be narrow, then one can describe it 
in a more simple way by a small number of $Z'$ production and decay characteristics. In this approach
it is only needed to set the $Z'$ mass, the production
cross-section, and the total and partial decay widths.
Being quite simple, such a scheme at the same time could give 
estimations of $Z'$ couplings to the SM fields based on the current 
experimental data.

It is possible to calculate effects of $Z'$ boson in details for
each specific GUT model. Such model-dependent estimates are widely
presented in the literature  \cite{Erler:2009jh,aguila,Erler:2010arxiv,Corcella:models2013,PankovAndreev:models2012,PankovAndreev:models2013,Cao:models2012,Basso:models2012,Belyaev:models2013}.
Some set of popular $E_6$-based models and left-right models is usually
considered in this approach. However, probing the set one can still
miss the actual $Z'$ model. Therefore, it is useful to
complement the model-dependent $Z'$ searches by some kind of
model-independent analysis (e.g. as in Ref. \cite{Eboli:indep2012}). Lots of the usually considered models belong to the
models with the so-called Abelian $Z'$ boson. The Abelian $Z'$ is usually understood as an effective additional $U(1)$ gauge state
at energies of order of several TeVs, which obtains its heavy mass beyond the scope of the SM. Such kind of $Z'$ boson is characterized
by specific relations between its couplings to SM particles. The relations were derived in Refs.
\cite{Gulov:2000eh,Gulov:1999ry}. They cover models satisfying the following conditions:
1) only one heavy neutral vector boson could be recognized at energies of modern colliders, whereas other possible heavy bosons are decoupled at essentially larger mass scales; 2) the $Z'$ boson is decoupled at low energies and can be phenomenologically described by an
effective Lagrangian \cite{leike,Lang08,Rizzo06}; 3)
the underlying theory matches with either one- or two-Higgs-doublet standard model at low energies;
4) the SM gauge group is a subgroup of the gauge group of the underlying theory; 5) $Z'$ boson is described by an effective additional $U(1)$ gauge state at low energies. While the relations are not model-independent in the most broad sense, they can be referred as applicable to a wide set of specific models.
Among the popular models discussed in the literature, the left-right models and the $E_6$ models belong to this set. Nevertheless, it does not mean that the approach is designed and applicable only to those models. 

It follows from the mentioned relations that the Abelian $Z'$
couplings to the left-handed fermion currents within any SM
doublet are the same and that the absolute value of the $Z'$ couplings to
the axial-vector currents for all the massive SM fermions is universal (see Refs. \cite{GulovSkalozub:2009review,GulovSkalozub:2010ijmpa}
for details). The relations reduce significantly the number of unknown $Z'$ parameters and leave some of them arbitrary, therefore allowing analysis of experimental data complementary to the common model-dependent approach.
For instance, some Abelian $Z'$ hints were found in LEP data \cite{GulovSkalozub:2010ijmpa}. 

In Ref. \cite{GulovKozhushko:2011ijmpa} two different estimates both for the $Z'$ production cross section at the LHC and the $Z'$ decay width were presented. Those are the \textit{95\% CL estimate}, where all the $Z'$ coupling constants are varied in their 95\% confidence level (CL)
intervals derived by LEP data,
and the \textit{maximum-likelihood estimate}, where the $Z'$ coupling to axial-vector currents is set to
its mean value from experimental data,
$a/m_{Z'}\simeq\pm 0.14\mathrm{~TeV}^{-1}$, and the fermionic couplings $v_{f}$ are varied in their 95\% CL intervals. It was shown that in case of the maximum-likelihood estimate at $Z'$ masses up to 1.5 TeV the narrow-width approximation (NWA) is applicable, and therefore it is possible to calculate 
the $Z'$ contribution to the Drell-Yan cross section as $\sigma (pp \to Z') \times BR(Z' \to l^+l^-)$.
However, this is not the case for higher $Z'$ mass region, namely, at $m_{Z'}$ $\sim$ 2-3 TeV, which has recently been explored by the LHC collaborations \cite{ATLAS:2013NW,CMS:2013NW}.
In this region the NWA condition $\Gamma_{Z'}^2/m_{Z'}^2 \ll 1$ for the {\it maximum-likelihood estimate} is not met.
The reason is that even in case of this very optimistic scenario the intervals for the vector couplings are still too wide.
Since the LHC collaborations present their results calculated in the NWA, then to be able to obtain estimates of the $Z'$ contribution to the Drell-Yan process at the LHC and compare them with the currently available data, we need to change our estimation strategy. 

In our present investigation we use the relations for the Abelian $Z'$ couplings to estimate
the Abelian $Z'$ production in the Drell-Yan process. We compare the
obtained cross section to the current LHC bounds. This allows
us to constrain $Z'$ couplings. It also shows how far the LHC will
potentially advance the $Z'$ searches compared to the LEP.

In Section 2 we provide all necessary information about the $Z'$ boson
and the used relations between couplings.
Section 3 contains some details regarding $Z'$ production and decay at the LHC.
In Section 4 our estimation strategy is presented.
In Section 5 we discuss the obtained results.

\section{$Z'$ parameterization}\label{sec:couplings}

In the present paper we use the following effective Lagrangian to describe $Z'$ couplings to the axial-vector and vector fermion currents:
\begin{eqnarray}\label{ZZplagr}
{\cal L}_{Z\bar{f}f}&=&\frac{1}{2} Z_\mu\bar{f}\gamma^\mu\left[
(v^\mathrm{SM}_{fZ}+\gamma^5 a^\mathrm{SM}_{fZ})\cos\theta_0
+(v_f+\gamma^5 a_f)\sin\theta_0 \right]f, \nonumber\\
{\cal L}_{Z'\bar{f}f}&=&\frac{1}{2} Z'_\mu\bar{f}\gamma^\mu\left[
(v_f+\gamma^5 a_f)\cos\theta_0
-(v^\mathrm{SM}_{fZ}+\gamma^5
a^\mathrm{SM}_{fZ})\sin\theta_0\right]f,
\end{eqnarray}
where $f$ is an arbitrary SM fermion state;
$a_f$ and $v_f$ are the $Z'$ couplings to the
axial-vector and vector fermion currents; $\theta_0$
is the $Z$--$Z'$ mixing angle;
$v^\mathrm{SM}_{fZ}$,
$a^\mathrm{SM}_{fZ}$ are the SM couplings of the $Z$-boson. 
The commonly considered $Z'$ gauge coupling $\tilde{g}$ is included into $a_f$ and $v_f$.

This popular parameterization follows from a number of natural conditions.
First of all, the $Z'$ interactions of renormalizable types are expected to
be dominant. The non-renormalizable
interactions that are generated at high energies due to radiation
corrections are suppressed by $1/m_{Z^\prime}$ (or by other heavier scales $1/\Lambda_i\ll
1/m_{Z^\prime}$) at low energies $\sim m_W$ and therefore they can be neglected.

We also assume the conditions listed in the Introduction in order to use the relations between the Abelian $Z^\prime$ couplings.
In particular, the SM gauge group ${\rm SU}(2)_L\times{\rm U}(1)_Y$ is considered as a
subgroup of the GUT group. In this case, a product of the SM subgroup
generators is a linear combination of these
generators. Consequently, all the structure constants
that connect the two SM gauge bosons with $Z^\prime$ have to be zero,
and at the tree-level $Z'$ interactions to the SM gauge fields are 
possible due to a $Z$--$Z'$ mixing only. 

To calculate the $Z'$ contribution to the Drell-Yan process, we
also need to parameterize the $Z'$ interactions with the SM scalar and
vector fields. The explicit Lagrangian
describing $Z'$ couplings to all the SM fields can be found in
Ref. \cite{GulovKozhushko:2011ijmpa}.

The parameters $a_f$, $v_f$, and $\theta_0$ could be obtained from
experimental data. In a particular model, one has some specific values
for some of them. If the model is unknown, all the parameters are potentially arbitrary numbers. 
If one assumes that the underlying extended model is renormalizable, then, 
as was shown in Refs. \cite{Gulov:2000eh,Gulov:1999ry}, there is a relation between these parameters:
\begin{equation} \label{grgav}
v_f - a_f= v_{f^*} - a_{f^*}, \qquad a_f = T_{3f}
\tilde{g}\tilde{Y}_\phi.
\end{equation}
Here, $f$ and $f^*$ are the components of the $SU(2)_L$ fermion
doublet ($l^* = \nu_l, \nu^* = l, q^*_u = q_d$, and $q^*_d = q_u$),
$T_{3f}$ is the third component of weak isospin (1/2 for ``up''-type fermions, -1/2 for ``down''-type fermions), and
$\tilde{g}\tilde{Y}_\phi$ determines $Z'$ couplings to the
SM scalar fields and the $Z$--$Z'$ mixing angle $\theta_0$ in (\ref{ZZplagr}),
%(see Ref. \citen{GulovKozhushko:2011ijmpa} for details).
which is expressed as:

\begin{eqnarray}\label{tan2theta}
\tan 2\theta_0 = \tilde{g}\tilde{Y}_\phi \frac{\sin \theta_W \cos\theta_W}{\sqrt{4 \pi \alpha_{\rm em}}} \frac{m_Z^2}{m_{Z'}^2}.
\end{eqnarray}

As it was argued in Refs.
\cite{GulovSkalozub:2009review,GulovSkalozub:2010ijmpa}, the
relations (\ref{grgav}) hold in a set of popular models with the Abelian $Z'$ boson based on
the ${\rm E}_6$ group (the so called LR, $\chi$-$\psi$ models). However, one could also think
about models beyond the commonly used list of models.

Let us note that the couplings of the Abelian $Z'$ to the axial-vector fermion
currents are described by a universal absolute value.
Therefore we introduce the notation
\begin{equation}\label{RGrel1}
a = a_d = a_{e^-} = -a_u = -a_{\nu}.
\end{equation}
From Eq. (\ref{grgav}) it follows, that this value $a$ is proportional
to the $Z'$ coupling to scalar fields. 
By substituting Eqs. (\ref{grgav}) and (\ref{RGrel1}) into Eq. (\ref{tan2theta}) we obtain
\begin{equation}\label{RGrel2}
\theta_0 \approx -2a\frac{\sin \theta_W \cos
\theta_W}{\sqrt{4 \pi \alpha_{\rm em}}} \frac{m_Z^2}{m_{Z'}^2}.
\end{equation}
Thus the $Z$--$Z'$ mixing angle $\theta_0$ is also determined by the axial-vector
coupling. For further calculations we use $\alpha_{\rm em} = 1/128.9$, $\sin^2 \theta_W = 0.2304$.

It can be seen from (\ref{grgav}), that for each fermion doublet
only one vector coupling is independent:
\begin{equation}\label{RGrel3}
v_{f_d} = v_{f_u} + 2 a.
\end{equation}
In total, the $Z'$ interactions with the SM particles
can be parameterized by seven
independent couplings: $a$, $v_u$, $v_c$,
$v_t$, $v_e$, $v_\mu$, $v_\tau$.

In Refs. \cite{GulovSkalozub:2009review,GulovSkalozub:2010ijmpa} 
the limits on $Z'$ couplings from the LEP I and
LEP II data were obtained. One can interpret those limits as
some hints of $Z'$ boson at 1-2$\sigma$ CL. Namely, the couplings
$a$ and $v_e$ show non-zero maximum-likelihood (ML) values. The
constraints on the axial-vector coupling $a$ come from the LEP I
data (through the mixing angle) and from the LEP II data on the
$e^+e^-\to\mu^+\mu^-,\tau^+\tau^-$ scattering. The corresponding ML
values lie very close to each other. In our estimates we use the value
\begin{equation}\label{MLV}
%{a}/m_{Z'}\simeq\pm 0.14\mathrm{TeV}^{-1}
a_\mathrm{ML}^2/m_{Z'}^2=1.97\times 10^{-2}\ \mathrm{TeV}^{-2}.
\end{equation}

The electron vector coupling $v_e$ is constrained at 95\% CL by the
$e^+e^-\to e^+ e^-$ data from LEP II (see discussion in Refs. 
\cite{GulovSkalozub:2009review, GulovKozhushko:2011ijmpa}):
\begin{equation}\label{CLI2}
6.07\times 10^{-2}\ \mathrm{TeV}^{-2}< {v}_e^2/m_{Z'}^2 <2.56\times 10^{-1}\ \mathrm{TeV}^{-2}.
\end{equation}
These constraints seem to be less stable, so we will use them only to compare our
final results avoiding taking them into account in calculations.

There are no significant constraints on the other $Z'$ coupling constants 
from the existing data. 

\section{$Z'$ production at the LHC}\label{sec:prod}

\begin{figure}
\centering{
%\begin{minipage}[b]{0.33\linewidth}
\psfig{file=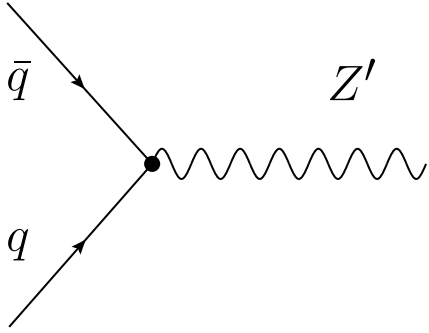,width=60mm}
 \caption{$Z'$ production at the parton level.}
 \label{fig:Prod}}
\end{figure}

At the LHC $Z'$ bosons are expected to be produced in
proton-proton collisions: $pp \to Z'$. At the parton level
this process is described by the $Z'$ production in the quark-antiquark
pair annihilation, $q\bar{q} \to Z'$ (Fig. \ref{fig:Prod}). The
cross-section of the $pp \to Z'$ process is 
obtained by integrating the partonic cross-section 
$\sigma_{q\bar{q}\to Z'}$ with the parton
distribution functions (PDFs):
\begin{eqnarray}
\sigma_{AB} &=&
\sum_{q,\bar{q}}\int_{0}^{1}dx_{q}\int_{0}^{1}dx_{\bar{q}} \, f_{q,A}
(x_q,\mu_R,\mu_F)f_{\bar{q},B}(x_{\bar{q}},\mu_R,\mu_F)
\nonumber\\
&&\times \sigma_{q\bar{q}\to Z'}(m_{Z'}, x_q k_A, x_{\bar{q}} k_B),
\end{eqnarray}
where $A$, $B$ mark the interacting hadrons (protons in our case)
with the four-momenta $k_A$, $k_B$; $f_{q,A}$ is the parton
distribution function for the parton $q$ in the hadron $A$ with
the momentum fraction $x_q$ at the renormalization 
scale $\mu_R$ and factorization scale $\mu_F$. 
We use the parton distribution functions provided by the MSTW PDF package \cite{mstw}.

The production cross-section includes quadratic
combinations of the $Z'$ couplings to quarks,
\begin{eqnarray}
\label{eq:factorcs}
\sigma_{AB} = a^2 \sigma_{a^2} + a v_u \sigma_{a v_u} + v_u^2 \sigma_{v_u^2} + a v_c \sigma_{a v_c} 
 + v_c^2 \sigma_{v_c^2} + a v_t \sigma_{a v_t} + v_t^2 \sigma_{v_t^2}.
\end{eqnarray}
Here we took into account relations (\ref{RGrel1})--(\ref{RGrel3}). 
The factors $\sigma$ on the right side of the previous equation
depend on $m_{Z'}$ and the beam energy. At energies above
2 TeV the factors $\sigma_{a v_c}$,
$\sigma_{v_c^2}$, 
$\sigma_{a v_t}$, and
$\sigma_{v_t^2}$ amount to less than 1\% of each of the factors
$\sigma_{a^2}$, $\sigma_{a v_u}$, and 
$\sigma_{v_u^2}$, and therefore we neglect
their contributions.

We take into account the 90\% CL uncertainty intervals for the parton
distributions provided in the MSTW PDF package, and also the 
uncertainties that arise from the renormalization and factorization scales variation: 
$\mu_R = \mu_F = \mu$, $ m_{Z'}/2 \leq \mu \leq 2 m_{Z'}$.

Both the $Z'$ production cross section and the uncertainties are calculated in the leading order in $\alpha_S$. 
The next-to-next-to-leading order cross section together with the corresponding uncertainties 
is obtained using the NNLO K-factor for the Drell-Yan process calculated in the Standard model:

\begin{eqnarray}
\label{eq:kfactor}
K = \frac{\sigma_{\mathrm{DY}}^{\mathrm{NNLO}}}{\sigma_{\mathrm{DY}}^{\mathrm{LO}}}.
\end{eqnarray}

It is calculated using the {FEWZ} software \cite{FEWZ}.
$K$ increases monotonically from $1.28 \pm 0.08$ to $1.30 \pm 0.06$, as $m_{Z'}$ varies from 2 TeV to 3 TeV.

Finally, the production cross-section reads:
\begin{eqnarray}
\label{eq:factorcsshort} \sigma_{pp \to Z'} & = & a^2 \sigma_{a^2}
+ a v_u \sigma_{a v_u} +
v_u^2 \sigma_{v_u^2} \pm \Delta \sigma^{\rm pdf + scale}, \nonumber\\
\Delta \sigma^{\rm pdf + scale} & = & a^2 \Delta \sigma^{\rm
pdf + scale}_{a^2} + a v_u \Delta \sigma^{\rm
pdf + scale}_{a v_u} + v_u^2 \Delta \sigma^{\rm
pdf + scale}_{v_u^2}.
\end{eqnarray}

\begin{figure}
\centering{\psfig{file=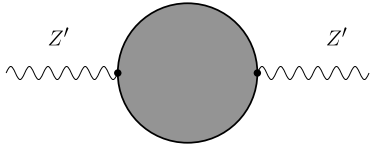,width=50mm}}
\caption{One-particle-irreducible correction to $Z' \to Z'$.} \label{fig:Width}
\end{figure}

The $Z'$ decay width $\Gamma_{Z'}$ is calculated using the
optical theorem:
\begin{eqnarray}
\Gamma_{Z'} = -\frac{{\rm Im}\,G(m_{Z'}^2)}{m_{Z'}}.
\end{eqnarray}
Here, $G(p^2)$ is the two-point one-particle-irreducible Green's
function, represented by the diagram in Fig. \ref{fig:Width}. The 
decay width $\Gamma_{Z'}$ is calculated at the one-loop level 
with the software packages
FeynArts, FormCalc, and LoopTools \cite{FeynArts,FormCalc}.
The Feynman diagrams with
internal $Z'$ lines and the Passarino-Veltman integrals of
type $A$ are real numbers and do not contribute to the decay width.
The remaining diagrams correspond to different $Z'$
decay channels. As a result, we obtain all the partial widths (and 
the branching ratios) corresponding to $Z'$ decays into certain pairs of
SM particles.

The partial width corresponding to $Z'$ decay
into a fermionic pair $f\bar{f}$ can be written in the following form:
\begin{equation}\label{eq:factorwidth}
\Gamma_{Z'\to \bar{f}f} = a_f^2 \Gamma_{a_f^2} + a_f v_f
\Gamma_{a_f v_f} + v_f^2 \Gamma_{v_f^2}.
\end{equation}
The factors $\Gamma_{a_f^2}$, $\Gamma_{a_f v_f}$, and
$\Gamma_{v_f^2}$ are proportional to $m_{Z'}$. 

\section{Estimation Scheme}

The main $Z'$ decay channels considered by ATLAS and CMS are 
dielectronic and dimuonic channels. The couplings that enter the corresponding cross sections
are $a$, $v_u$, and $v_e$ for the $pp\to Z' \to e^+ e^-$ case and
$a$, $v_u$, and $v_\mu$ for the $pp\to Z' \to \mu^+ \mu^-$ case.

Since $v_{\mu}$ was not constrained by the LEP data, we are going to study only the dielectron final state (also note that both these processes are similar at high energies). This allows us to estimate how the LHC data limits the $Z'$ couplings compared to the LEP results.

Let us present our estimation scheme.
Since there is a maximum-likelihood value for $a^2$ from LEP, $a_\mathrm{ML}^2/m_{Z'}^2=1.97\times 10^{-2}\mathrm{~TeV}^{-2}$, we can consider it
as our ``optimistic'' estimate. There is a ``pessimistic'' estimate
with $a^2 = 0$ for weakly-coupled $Z'$. To obtain a kind of an arbitrary estimate, we also consider $a^2 = a_\mathrm{ML}^2/4$.
Replacing the axial-vector coupling by these three estimates in
the $pp \to Z' \to e^+ e^-$ cross section, we can investigate possible $v_e$ and $v_u$ values taking onto account the LHC results on direct searches for $Z'$ resonances \cite{ATLAS:2013NW,CMS:2013NW}.

First, we need to determine the region of couplings in which the NWA is applicable. The criterion is $\Gamma_{Z'}^2 / m_{Z'}^2 \ll 1$. We set it to 

\begin{equation}
\label{eq:NWAcriterium}
\Gamma_{Z'}^2 / m_{Z'}^2 \leq 0.01
\end{equation}

To obtain the widest possible region for $v_e$ and $v_u$, we set the rest of vector couplings to the values at which the corresponding partial widths are minimal. From Eq. (\ref{eq:factorwidth}) and taking into account relations (\ref{RGrel1}), (\ref{RGrel3}) those values are:

\begin{equation}\label{eq:vectormin}
v_f = - \frac{a \Gamma_{a v_f}}{2\Gamma_{v_f^2}} \approx \pm 1\times a, \quad f \neq e^{-}, u,
\end{equation}
where the plus sign is for leptonic couplings, and the minus sign is for quark couplings.

Our next step is to investigate how the currently available LHC data constrains the values of $v_e$ and $v_u$. Both the ATLAS \cite{ATLAS:2013NW} and CMS \cite{CMS:2013NW} results indicate that the lower bounds for the $Z'$ mass lie between 2~TeV and 3~TeV. Therefore, we shall derive our constraints for those two values. The $pp \to Z' \to e^+ e^-$ cross section is calculated as

\begin{equation}\label{eq:NWAcs}
\sigma_{NWA} = \sigma_{pp \to Z'} \times BR(Z' \to e^+ e^-) = \sigma_{pp \to Z'} \times \frac{\Gamma_{Z' \to e^+ e^-}}{\Gamma_{Z'}}.
\end{equation}

We compare this cross section to the experimental upper bounds presented in Refs. \cite{ATLAS:2013NW} and \cite{CMS:2013NW} for $pp$ collisions at $\sqrt{S} = 8$ TeV. At the considered $Z'$ mass values it is always possible to choose such $v_f$ ($f \neq e^{-}, u$) values, that
correspond to the upper bound of the $Z'$ decay width in Eq. (\ref{eq:NWAcriterium}). Therefore, both for $m_{Z'} = 2$ TeV and $m_{Z'} = 3$ TeV we set $\Gamma_{Z'}$ to $0.1 \times m_{Z'}$. This will allow us to obtain widest possible LHC-driven intervals for $v_e$ and $v_u$.

\section{Results and Discussion \label{sec:discussion}}

\begin{figure}[t!]
\begin{minipage}[][][b]{0.32\linewidth}
\includegraphics[width=\linewidth]{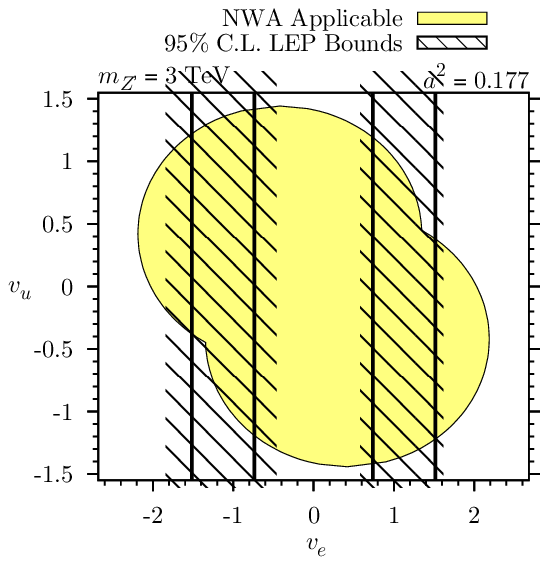}
\end{minipage}
\hfill
\begin{minipage}[][][b]{0.32\linewidth}
\includegraphics[width=\linewidth]{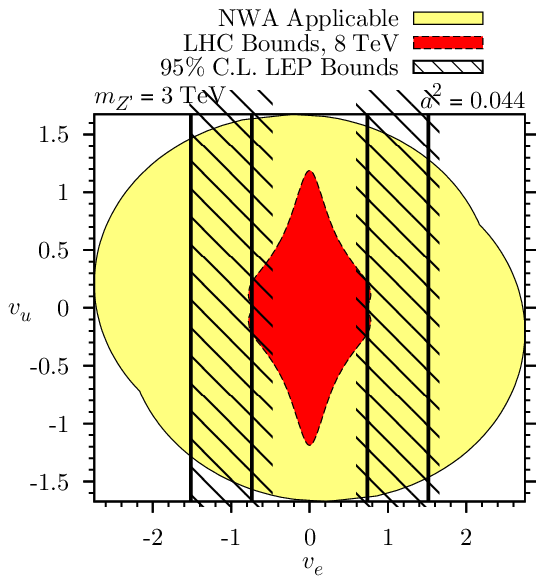}
\end{minipage}
\hfill
\begin{minipage}[][][b]{0.32\linewidth}
\includegraphics[width=\linewidth]{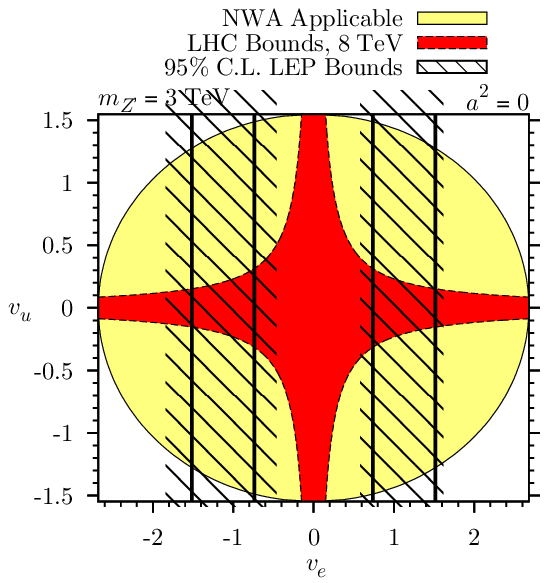}
\end{minipage}
\vfill
\begin{minipage}[][][b]{0.32\linewidth}
\includegraphics[width=\linewidth]{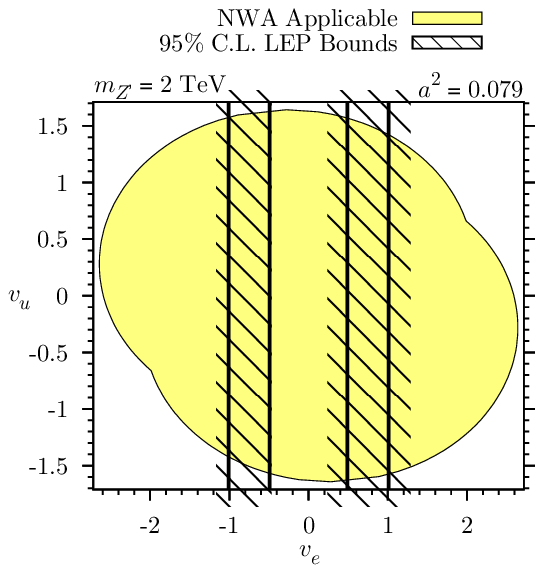}
\end{minipage}
\hfill
\begin{minipage}[][][b]{0.32\linewidth}
\includegraphics[width=\linewidth]{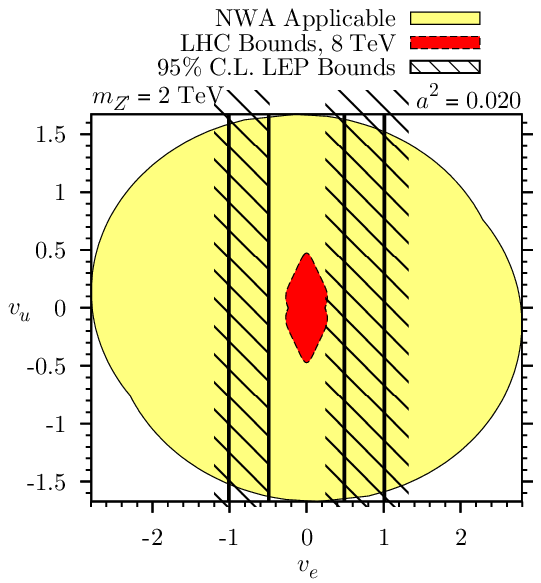}
\end{minipage}
\hfill
\begin{minipage}[][][b]{0.32\linewidth}
\includegraphics[width=\linewidth]{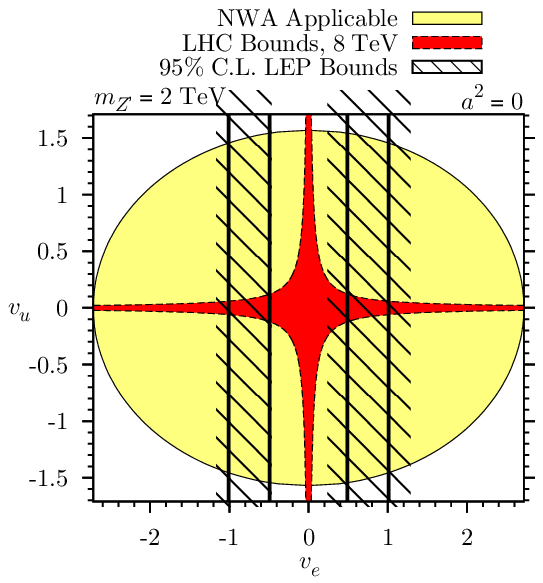}
\end{minipage}
 \caption{The $v_u$-vs-$v_e$ planes. First row: $m_{Z'} = 3$~TeV; second row: $m_{Z'} = 2$~TeV. The first column is for the ``optimistic'' estimation scheme, the second column is for the intermediate estimation, and on the plots of the third column the ``pessimistic'' scheme is presented. The light-gray (yellow) areas represent the coupling values for which the narrow-width approximation is applicable (i.e. $\Gamma_{Z'}/m_{Z'} \leq 0.1$). The hatched areas are for the 95\% CL bounds on the $v_e$ coupling (from LEP II data). The dark-gray (red) area represents the $v_u$ and $v_e$ values allowed by the LHC.}
 \label{fig:LHCestimations}
\end{figure}

The constraints are shown in Fig. \ref{fig:LHCestimations} on the $v_u$-vs-$v_e$ planes. 
We present the areas of $v_u$ and $v_e$ values for which the NWA is applicable. For the ``optimistic'' estimation we use two possible values for the axial-vector coupling: $a/m_{Z'}\simeq\pm 0.14\mathrm{~TeV}^{-1}$. Also the LEP bounds for $v_e$ are shown for comparison. 

The ATLAS collaboration \cite{ATLAS:2013NW} reports upper bounds for $\sigma_{pp \to Z' \to e^+ e^-}$ at $1.5 \times 10^{-3}$ pb for $m_{Z'} = 2$ TeV and $2.5 \times 10^{-3}$ pb for $m_{Z'} = 3$ TeV. The ``optimistic'' estimation for $\sigma_{NWA}$ lies higher than these values, therefore, the LEP maximum-likelihood value $a_\mathrm{ML}^2/m_{Z'}^2 = 1.97 \times 10^{-2} \mathrm{~TeV}^{-2}$ is discouraged by the LHC results for $m_{Z'}$ from 2 TeV to 3 TeV. The LEP maximum-likelihood value is consistent with the LHC results for $Z'$ masses below 700~GeV. The region for the ``pessimistic'' and ``intermediate'' estimations overlaps with the LHC values, therefore allowing for non-zero upper bounds for the vector couplings. The regions of those are also plotted in Fig. \ref{fig:LHCestimations}. This indicates that the maximum-likelihood LEP value $a_\mathrm{ML}^2$ is disfavored approximately by one order of magnitude. These two estimates represent a $Z'$ with small coupling to the axial-vector currents.

From the plots for the presented estimation schemes we can see that the LHC may limit the vector couplings to $v_u^2/m_{Z'}^2 \leq 10^{-2} .. 10^{-3}~\mathrm{TeV}^{-2}$, $v_e^2/m_{Z'}^2 \leq 10^{-1} .. 10^{-2}~\mathrm{TeV}^{-2}$, which is one order lower than the LEP limits. In the considered $Z'$ mass region these values are larger than the respective couplings of the SM $Z$ boson ($10^{-3}$ for $v_e^\mathrm{SM}$ and $2 \times 10^{-2}$ for $v_u^\mathrm{SM}$), but at low energies the $Z'$ interactions are strongly suppressed by $1/m_{Z'}$. Also it has to be noted, that the renormalization-group relations (\ref{RGrel1}) used in this paper are not applicable for the standard-model $Z$ boson because of different group structure.

It is interesting to calculate the $Z$--$Z'$ mixing angle value based on our estimations. Current LEP-driven upper limits for $\theta_0$ in different $Z'$ models are of order of $10^{-3}$ (see Table IV in Ref. \cite{Lang08}). For our ``optimistic'' estimate Eq. (\ref{RGrel2}) gives $1.6 \times 10^{-3}$ for the $\theta_0$ value considering $m_{Z'} = 2$ TeV. As it was noted, this value is all but ruled out by the LHC data, so for Ableian $Z'$ models one may expect $\theta_0$ less than (a few)$\times 10^{-4}$.

To obtain more strict bounds, one has to take into account the contributions from the remaining fermions and consider the differential cross-section, rather than working in the narrow-width approximation. Nevertheless,  the two presented estimates, being rough, still allow to see, how far it is possible to advance both the direct and indirect $Z'$ searches compared to the LEP results.

%\section{Bibliography}


\begin{thebibliography}{99}
\bibitem{leike} A. Leike, \Journal{\PRP}{317}{143}{1999}{The phenomenology of extra neutral gauge bosons}.
%
\bibitem{Lang08} P. Langacker, \Journal{\RMP}{81}{1199-1228}{2008}{The Physics of Heavy $Z'$ Gauge Bosons}.
%\eprint{arXiv:0801.1345 [hep-ph]}{}.
%
\bibitem{Rizzo06} T. Rizzo, $Z'$ Phenomenology and the LHC, in \textit{Colliders and neutrinos: The window into physics beyond the standard model; Proc of Summer School TASI 2006, Boulder, USA, June 4-30, 2006}, eds. S.~Dawson and R.N.~Mohapatra, p.537-575, \eprint{hep-ph/0610104}..
%
\bibitem{ATLAS:2013NW} N. Hod (on behalf of the ATLAS collaboration), Search for heavy resonances, and resonant diboson production with the ATLAS detector, \textit{Proceedings of Hadron Collider Physics Symposium 2012 (HCP 2012), Kyoto, Japan, November 12-16, 2012}, eds. M.~Ishino, K.~Nagano, S.~Asai,  \textit{EPJ Web Conf.} \textbf{49}, 15004 (2013), \eprint{arXiv:1303.4287 [hep-ex]}.

\bibitem{CMS:2013NW} CMS Collaboration, \Journal{\PLB}{720}{63}{2013}, \eprint{arXiv:1212.6175 [hep-ex]}{Search for heavy narrow dilepton resonances in $pp$ collisions at $\sqrt{s} = 7$ TeV and $\sqrt{s} = 8$ TeV}.
%
\bibitem{Boos_ea_1}E. Boos, V. Bunichev, L. Dudko and M. Perfilov,
\Journal{\PLB}{655}{245}{2007}.

\bibitem{Boos_ea_2}E.E. Boos, M.A. Perfilov, M.N. Smolyakov and I.P. Volobuev,
\Journal{Theor.Math.Phys.}{170}{90-96}{2012}.

\bibitem{PankovTsytrinov:2009prd} P. Osland, A. A. Pankov, N. Paver and A. V. Tsytrinov \Journal{\PRD}{79}{115021}{2009}{Spin and model identification of $Z'$ bosons at the LHC}.
%
\bibitem{GulovKozh:2012obs} A. Gulov and A. Kozhushko, \eprint{arXiv:1209.5022 [hep-ph]}{Two-parametric model-independent observables for \boldmath $Z'$ searching at the Tevatron}.
%
\bibitem{Erler:2009jh} J.~Erler, P.~Langacker, S.~Munir and E.\,R. Pena,
\Journal{JHEP}{08}{017}{2009}{Improved Constraints on $Z'$ Bosons from Electroweak Precision Data}.

\bibitem{aguila} F.~del Aguila, J.~de Blas and M.~Perez-Victoria, \textit{JHEP} \textbf{1009}, 033 (2010), \eprint{arXiv:1005.3998}.
%
\bibitem{Erler:2010arxiv} J. Erler, P. Langacker, S. Munir and E. Rojas,
\eprint{arXiv:1010.3097v1 [hep-ph]}{$Z'$ Searches: From Tevatron to LHC}.
%
\bibitem{Corcella:models2013} G. Corcella, \textit{EPJ Web Conf.} \textbf{60}, 18011 (2013), \eprint{arXiv:1307.1040 [hep-ph]}.
%
\bibitem{PankovAndreev:models2012} V. V. Andreev, G. Moortgat-Pick, P. Osland, A. A. Pankov and N. Paver, \eprint{arXiv:1205.0866 [hep-ph]}{Discriminating $Z'$ from anomalous trilinear gauge coupling signatures in $e^+ e^- \to W^+ W^-$ at ILC with polarized beams}
%
\bibitem{PankovAndreev:models2013} V. V. Andreev, G. Moortgat-Pick, P. Osland, A. A. Pankov and N. Paver, \textit{\EPJC} \textbf{72}, 2147 (2012), \eprint{arXiv:1205.0866 [hep-ph]}.

\bibitem{Cao:models2012} Q.-H. Cao, Z. Li, J.-H. Yu and C.-P. Yuan, \Journal{\PRD}{86}{095010}{2012}{Discovery and Identification of $W'$ and $Z'$ in $SU(2)xSU(2)xU(1)$ Models at the LHC}.

\bibitem{Basso:models2012} L. Basso, K. Mimasu and S. Moretti, \Journal{\JHEP}{1211}{060}{2012}{Non-exotic $Z'$ signals in $l^+ l^-$,$b\bar{b}$ and $\bar{t}{t}$ final states at the LHC}.

\bibitem{Belyaev:models2013} E. Accomando, D. Becciolini, A. Belyaev, S. Moretti and C. Shepherd-Themistocleous, \textit{JHEP} \textbf{10}, 153 (2013), \eprint{arXiv:1304.6700 [hep-ph]}.

\bibitem{Eboli:indep2012} O.J.P. Eboli, J. Gonzalez-Fraile and M.C. Gonzalez-Garcia, \Journal{\PRD}{85}{055019}{2012}{Present Bounds on New Neutral Vector Resonances from Electroweak Gauge Boson Pair Production at the LHC}.
%
\bibitem{Gulov:2000eh} A.\,V. Gulov and V.\,V. Skalozub,
\Journal{\EPJC}{17}{685}{2000}{Renormalizability and model independent description of $Z'$ signals at low-energies}.
%
\bibitem{Gulov:1999ry} A.\,V. Gulov and V.\,V. Skalozub,
\Journal{\PRD}{61}{055007}{2000}{Renormalizability and the model independent observables for Abelian $Z'$ search}.
%
\bibitem{GulovSkalozub:2009review} A.\,V. Gulov and V.\,V. Skalozub, \eprint{arXiv:0905.2596v2 [hep-ph]}{Model independent search for $Z'$ boson signals}.
%
\bibitem{GulovSkalozub:2010ijmpa} A.\,V. Gulov and V.\,V. Skalozub,
\Journal{\IJMPA}{25}{5787-5815}{2010}{Fitting of $Z'$ parameters}.
%
\bibitem{GulovKozhushko:2011ijmpa} A.\,V. Gulov and A.\,A. Kozhushko, \Journal{\IJMPA}{26}{4083-4100}{2011}{Model-independent estimates for the abelian $Z'$ boson at modern hadron colliders}.
%
\bibitem{mstw} A.D. Martin, W.J. Stirling, R.S. Thorne and G. Watt, \Journal{\EPJC}{63}{189}{2009}{Parton distributions for the LHC}; \Journal{\ibid}{64}{653}{2009}; http://projects.hepforge.org/mstwpdf/.
%
\bibitem{FEWZ} R. Gavin, Y. Li, F. Petriello and S. Quackenbush, \Journal{\CPC}{182}{2388-2403}{2011}; http://gate.hep.anl.gov/fpetriello/FEWZ.html.
%
\bibitem{FeynArts} T. Hahn, \Journal{\CPC}{140}{418}{2001}{Generating Feynman diagrams and amplitudes with FeynArts 3}; http://www.feynarts.de/.
%
\bibitem{FormCalc} T. Hahn and M. Perez-Victoria, \Journal{\CPC}{118}{153}{1999}{Automatized one loop calculations in four-dimensions and D-dimensions}; http://www.feynarts.de/formcalc/, http://www.feynarts.de/looptools/.

\end{thebibliography}
\end{document}